\documentclass[pra,superscriptaddress,aps,twocolumn,10pt,showkeys]{revtex4-1}
\usepackage{graphics,graphicx,epsfig,color,latexsym,float,amsmath,bm}
\usepackage[colorlinks,citecolor=blue,linkcolor=blue,urlcolor=blue]{hyperref}
\usepackage[version=4]{mhchem}
\usepackage{wrapfig}
\usepackage{tikz}
\usetikzlibrary{arrows,chains,decorations.pathreplacing,decorations.markings}
\usepackage{amsfonts,amssymb,bm}

\graphicspath{{figures/}}

\usepackage{ulem}
\usepackage{lipsum}
\usepackage{placeins}

\usepackage{tikz}
\usetikzlibrary{quantikz}

\begin{document}
\title{Tensor train optimization of parametrized quantum circuits}

\author{Georgii Paradezhenko}
\affiliation{Skolkovo Institute of Science and Technology, Moscow 121205, Russia}
\author{Anastasiia Pervishko}
\affiliation{Skolkovo Institute of Science and Technology, Moscow 121205, Russia}
\affiliation{Leonhard Euler International Mathematical Institute, Saint Petersburg 199034, Russia}
\author{Dmitry Yudin}
\affiliation{Skolkovo Institute of Science and Technology, Moscow 121205, Russia}
\affiliation{Leonhard Euler International Mathematical Institute, Saint Petersburg 199034, Russia}

\date{\today}

\begin{abstract}
We examine a particular realization of derivative-free method as implemented on tensor train based optimization to the variational quantum eigensolver. As an example, we consider parametrized quantum circuits composed of a low-depth hardware-efficient ansatz and Hamiltonian variational ansatz for addressing the ground state of the transverse field Ising model. We further make a comparison with gradient-based optimization techniques and discuss on the advantage of using tensor train based optimization, especially in the presence of noise.
\end{abstract}

\maketitle

{\it Motivation.---}Significant progress towards stable operation of multi-qubit quantum systems with relatively short decoherence times allows nowadays to address simple optimization tasks~\cite{Harrigan2021,Ebadi2022,Yarkoni2022,Nguen2023}. The wider use of these noisy intermediate-scale quantum processors is hampered by the noise inevitably present in quantum gates, which severely limits the possible depth of a quantum circuit. The problem can nevertheless be partially relaxed in the approach of variational quantum computing, widely accepted as the most viable way to achieve quantum supremacy~\cite{McClean2016,Babbush2016,Yang2017,Paesani2017,Li2017,Dunjko2018,Preskill2018,Hempel2018,Colless2018,Santagati2018,Babbush2018,Kivlichan2018,Moll2018,LaRose2019,Schuld2019,Huggins2019,Gyongyosi2019,Cross2019,McArdle2019,Lee2019,Yuan2019,Zhu2019,Wang2019,Liu2019,Carolan2020,Uvarov2020,McArdle2020,Schuld2020,Endo2020,Lubasch2020,Kardashin2020,Cerezo2021,Kardashin2021,Biamonte2021,Monroe2021,Alexeev2021,Harrow2021,Skolik2021}. In variational quantum algorithms, a parameterized quantum circuit is trained by means of standard optimization algorithms in the same way as in machine learning. Despite the significant progress in experimental demonstration of this method, there is still a lack of theoretical understanding of the efficiency of using such algorithms in general. 

As a rule, in most variational quantum algorithms, with the variational quantum eigensolver (VQE) being the most notable example, one looks for the ground state of a given interacting quantum system~\cite{PMS14}. In this case, a quantum processor is used to prepare a family of probe states $\vert\psi(\bm{\theta})\rangle$ as implemented by a quantum circuit parametrized by $\bm{\theta}=(\theta_1,\theta_2,\ldots)$, as well as to estimate the energy for that family of state representing thus a multi-parameter cost function. At the next step, by virtue of standard optimization methods on a classical computer one minimizes the cost function to determine the optimal parameters of the quantum circuit that approximate the ground state of a given Hamiltonian $H$, {\it i.e.}, $\bm{\theta}^\ast=\mathrm{arg}\,\mathrm{min}\,\langle\psi(\bm{\theta})\vert H\vert\psi(\bm{\theta})\rangle$. The main advantage of this methodology is in the fact that one does not need to design a deep quantum circuit~\cite{Malley2016,KMT17,McClean2018}. Instead, the efficiency of the ground state approximation with variational quantum algorithms is extremely sensitive to the family of probe states. Choosing too shallow quantum circuits may turn out to be insufficient to approximate the desired state; in turn, by choosing a deep quantum circuit may lead to difficulties in optimizing the multi-parameter cost function. In particular, the optimization subroutine might be potentially plagued by the presence of barren plateau~\cite{McClean2018}.

Remarkably, a quantum circuit can be associated with a tensor network state. Extracting ground and low excited states as based on variational optimization with tensor networks has lately seen a surge of interest~\cite{Verstraete2008,Schollwock2011,Orus2014,Orus2019}. Note this approach is dated back to the original proposal of density matrix renormalization group by White~\cite{White1992}. If the state is specified by a tensor network one can easily evaluate the expectation values of operators to assess the properties of the ground state~\cite{Schollwock2005}. Analogously, thermal sates can be approached within the  imaginary time evolution of the density matrix in matrix product operators form~\cite{Pirvu2010}. In our work, we propose an opposite use of tensor networks, namely we apply the technique known as Tensor Train Optimizer (TTOpt)~\cite{SCS22} to explore the energy landscape as specified by the parametrized cost function in VQE. The methodology was devised to solve optimization tasks in a derivative-free manner. 

{\it Optimization strategy.---}In the typical VQE settings, we aim to estimate the expectation value such as $A(\bm{\theta}) = \bra{\psi(\bm{\theta})} A \ket{\psi(\bm{\theta})}$, where $A$ is a arbitrary Hermitian operator that is more often chosen in the form of the Hamiltonian~\cite{PMS14}. This is done with an available quantum processor or by means of numerical simulations. The probe state $\vert\psi(\bm{\theta})\rangle$ is prepared by applying parameterized circuits to the reference state $\mathcal{U}(\bm{\theta})\vert 0\rangle^{\otimes n}$ with $n$ specifying the number of input qubits. Once we evaluate $A(\bm{\theta})$ the classical subroutine in the form of optimization has to be invoked to find out the update to $\bm{\theta}$. If the given accuracy is reached the algorithm stops, otherwise it proceeds till the predefined amount of iterations. In most of the cases, one performs this classical subroutine upon applying different variations of gradient descent, {\it e.g.}, stochastic gradient descent. For a fixed learning rate $\eta$ that is known to be the hyperparameter of the model the new value of $\theta$ is determined from $\bm{\theta}_{i+1}=\bm{\theta}_i-\eta\nabla_\theta A(\bm{\theta}_i)$. 

It is however believed that derivative-free methodology to be more noise-resilient, including but not limited to Nelder-Mead algorithm and Powell’s conjugate direction method. An advantage of using derivative-free based optimizers heavily relies on their better upper bounds on the convergence~\cite{Harrow2021}. Noise mitigation can be also achieved basing on the simultaneous perturbation stochastic algorithm that evaluates a gradient using the methodology of finite difference in a random direction. Despite the larger variance of the gradient as compared to the standard routine it provides rather low cost of each individual iteration that requires an estimate of the cost function in just two points. We herein propose a derivative-free optimization technique based on TTOpt. 

{\it Model Hamiltonian and parameterized circuit.---}In the following, we are to make use of the transverse field Ising model (TFIM) as a testbed. The TFIM corresponds to a one-dimensional quantum spin chain with pairwise interaction between nearest neighbors and is described by the Hamiltonian,
\begin{equation}\label{TFIM}
    H = -\sum_{i=1}^{n-1} Z_i Z_{i+1} - h \sum_{i=1}^n X_i=H_1+H_2.
\end{equation}
We adopt herein the notation from quantum information with $X$ and $Z$ being the corresponding Pauli matrices. In Eq.~\eqref{TFIM}, $h$ quantifies the strength of the transverse field relative to the exchange coupling strength. We consider the chain of $n$ spins in the vicinity of quantum phase transition $h = 1$, provided open boundary conditions are imposed.
We choose open boundary conditions in the TFIM, because the VQE algorithm in this case faces the convergence issues when utilizing shallow circuits~\cite{LJG21}.

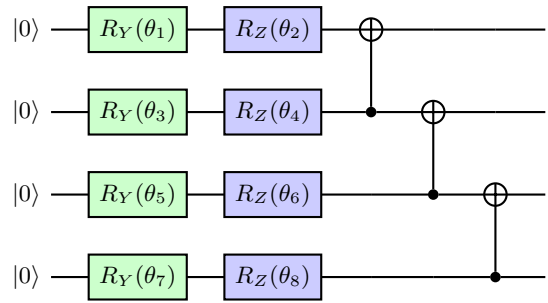
\begin{figure}[t]
\centering
\begin{quantikz}
\lstick{\ket{0}} & \gate[style={fill=green!20}]{R_{Y}(\theta_1)} & \gate[style={fill=blue!20}]{R_{Z}(\theta_2)} & \targ{} & \qw & \qw & \qw & \\
\lstick{\ket{0}} & \gate[style={fill=green!20}]{R_{Y}(\theta_3)} & \gate[style={fill=blue!20}]{R_{Z}(\theta_4)} & \ctrl{-1} & \targ{} & \qw & \qw & \\
\lstick{\ket{0}} & \gate[style={fill=green!20}]{R_{Y}(\theta_5)} & \gate[style={fill=blue!20}]{R_{Z}(\theta_6)} & \qw & \ctrl{-1} & \targ{} & \qw & \\
\lstick{\ket{0}} & \gate[style={fill=green!20}]{R_{Y}(\theta_7)} & \gate[style={fill=blue!20}]{R_{Z}(\theta_8)} & \qw & \qw & \ctrl{-1} & \qw &   \\
 \end{quantikz}
\caption{A one-layer hardware efficient ansatz (HEA) for the case of $n = 4$ qubits. The single-qubit rotation block is constituted by a sequence of $X$ and $Z$ rotations, where $R_{X}(\theta) = e^{-i\theta X/2}$ and $R_{Z}(\theta) = e^{-i\theta Z/2}$. The entangling block is represented by CNOT gates. The total number of parameters needed to parametrize a single layer of the $n$-qubits HEA is $2n$.}
\label{fig:hea}
\end{figure}

The proper choice of the parametrized circuit as provided by the unitary $\mathcal{U} (\bm{\theta})$ can be a challenging task for general Hamiltonians~\cite{SJA19}. Generally, there are two different approaches for constructing a quantum variational ansatz. The first one relies on the use of the hardware efficient ansatz (HEA) that can be directly implemented with existing topology of NISQ hardware~\cite{KMT17}. Specifically, the HEA circuit consists of $L$ layers of single-qubit rotations followed by entangling two-qubit operations (see Fig.~\ref{fig:hea}). The second approach is more physically motivated and based on a systematic approximation to the exact electronic wave function of the original problem. As an example one can think of the coupled-cluster~\cite{RBM19}, low-depth circuit ansatz~\cite{DD19}, and the Hamiltonian variational ansatz (HVA)~\cite{LJG21}. In our VQE simulations, we prepare probe states using the HEA and HVA.

\begin{figure}[ht!]
\centering
\begin{quantikz}
\lstick{\ket{0}} & \gate{H} & \gate[style={fill=blue!20},wires=2]{R_{ZZ}(\theta_1)} & \qw & \gate[style={fill=red!20}]{R_X(\theta_2)} & \qw \\
\lstick{\ket{0}} & \gate{H} & & \gate[style={fill=blue!20},wires=2]{R_{ZZ}(\theta_1)} & \gate[style={fill=red!20}]{R_X(\theta_2)} & \qw   \\
\lstick{\ket{0}} & \gate{H} & \gate[style={fill=blue!20},wires=2]{R_{ZZ}(\theta_1)} & & \gate[style={fill=red!20}]{R_X(\theta_2)} & \qw \\
\lstick{\ket{0}} & \gate{H} & & \qw & \gate[style={fill=red!20}]{R_X(\theta_2)} & \qw 
 \end{quantikz}
\caption{A one-layer Hamiltonian variational ansatz (HVA) for $n = 4$ qubits. The entangling block couples neighboring qubits in a pairwise manner. A two-qubit gate on a pair of qubits $i$ and $j$ is identified by the operator $R_{ZZ}(\theta)=e^{-i\theta Z_iZ_j/2}$, while $R_X$-gate on qubit $i$ represents $X$ rotation. The first layer of Hadamard gates, marked by $H$, prepares superposition of $\ket{0}$ and $\ket{1}$. A single layer of the $n$-qubits HVA consists of 2 parameters only, namely $\theta_1$ and $\theta_2$.}
\label{fig:hva}
\end{figure}
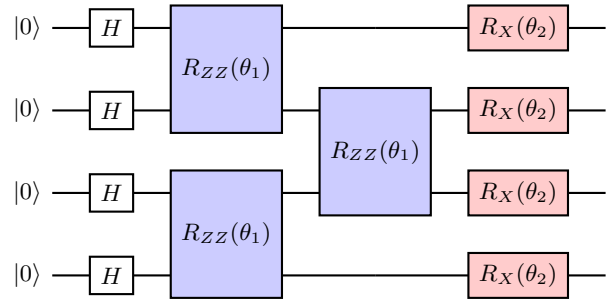

Suppose the Hamiltonian $H = \sum c_k H_k$ is decomposed into $K$ Pauli strings $H_k$
with $c_k$ specifying the real-valued coefficients. In this case, the HVA reads
\begin{equation}\label{HVA}
    \ket{\psi(\bm{\theta})} = \prod_{l=1}^L \left( \prod_{k=1}^K e^{-i \theta_{l,k} H_k} \right) \ket{\psi_0}=\mathcal{U}(\bm{\theta})\ket{\psi_0},
\end{equation}
where $\ket{\psi_0}$ is any quantum state that can be easily prepared. Thus defined ansatz \eqref{HVA} encompasses $L$ layers, so that each layer is constituted of $K$ blocks that correspond to the Hamiltonian decomposition in terms of Pauli strings and the total number of parameters used is $L\times K$. With the Hamiltonian of TFIM \eqref{TFIM} the depth-$L$ HVA circuit is defined by
\begin{equation}\label{HVA-TFIM}
  \mathcal{U}(\bm{\theta})
  = \prod_{l=1}^L e^{-i \theta_{l,1} H_1/2} \, e^{-i \theta_{l,2}H_2/2},
\end{equation}
which is shown in Fig.~\ref{fig:hva} in case of $n = 4$ qubits for $L = 1$ layer. Note that we choose $\ket{\psi_0}$ in Eq.~\eqref{HVA} to be the ground state of $H_2$, {\it i.e.}, $\ket{\psi_0}  =\ket{+}^{\otimes n}$.

It is worth mentioning that the HVA is claimed to be a powerful ansatz for finding the ground state of $H$, as it encodes the adiabatic evolution of the Hamiltonian~\cite{HWG19}. As demonstrated in Ref.~\cite{WZS20}, the HVA exhibits favorable structural properties such as mild or entirely absent barren plateaus for the TFIM and a restricted state space that eases their optimization in comparison to the HEA. The optimization landscape of the HVA becomes almost trap free, {\it i.e.}, there are no suboptimal minima, when the ansatz is overparameterized. Moreover, the HVA allows to significantly reduce the size of parameter space in comparison to the HEA, which is suitable for the tensor train based optimization.

When implementing the VQE on real quantum computers, one faces noise issues. The effect of noise makes the VQE to provide the results that can gradually deviate from the ground state energy. In practice, we want to make a quantum circuit as short as possible since the noise accumulates with increasing depth $L$ of a circuit. Although there are different gate local noise models including amplitude damping, dephasing, and depolarizing errors~\cite{NC10}, we consider only the depolarizing one since it accumulates with $L$ and significantly affects the VQE results in comparison to other two quantum errors~\cite{ZWC21}. We apply this quantum error for all one- and two-qubit operators involved in the ansatz circuits.  

For a quantum system of $n$ qubits, a completely positive trace-preserving map that describes the depolarizing error channel is specified by 
\begin{equation}\label{eq:noise}
  E(\rho) = (1-\lambda)\rho + \frac{\lambda}{2^n} I, 
\end{equation}
and the only parameter $0\leq \lambda\leq 2^{2n}/(2^{2n}-1)$. Clearly, Eq.~\eqref{eq:noise} transforms the quantum state $\rho$ into the linear combination of $\rho$ and maximally mixed state, with the latter being proportional to the identity matrix $I$~\cite{NC10}.

{\it On tensor train based optimization.---}In the VQE simulations, our computational task is to opitimize a real-valued cost function $A(\bm{\theta}) = \bra{\psi(\bm{\theta})} A \ket{\psi(\bm{\theta})}$, where $\bm{\theta} = (\theta_1,\theta_2,\ldots,\theta_d) \in \mathbb{R}^d$ is an $d$-dimensional vector of the variational parameters. Since all parameters are bounded, $\theta_i \in [0,2\pi]$, one can introduce the $d$-dimensional discretization grid for $\bm{\theta}$ on $\Omega = [0,2\pi]^{\times d}$ with $N$ nodes per each dimension. Now we can regard the function $A(\bm{\theta})$ as an implicit $d$-dimensional tensor $A_{i_1 i_2\ldots i_d}$, where the superscript $(i_1,\,i_2,\ldots,\,i_d)$ labels the grid nodes and $1\leq i_k\leq N$, $k=1,\,2,\ldots,\,d$. Determining the minimum of the function $A(\bm{\theta})$ for $\bm{\theta} \in \Omega$ translates into finding the maximal element of the tensor $-A_{i_1i_2\ldots i_d}$ in the discrete setting. Note that the total number of elements of the tensor $N^d$ scales up exponentially with $d$, making it being hardly evaluated or stored for sufficiently large $d$. In what follows, we show how to relax this issue in the spirit of TTOpt~\cite{SCS22}.

To grasp the very idea of the method consider an implicitly given matrix $A = (A_{i_1i_2}) \in \mathbb{R}^{N_1\times N_2}$, in other words a two-dimensional tensor. The method is based on the low-rank $R\ll \min\lbrace N_1, N_2 \rbrace$ cross approximation $\tilde{A}$ of the matrix $A$~\cite{CC10,ACC21}, 
\begin{equation}\label{cross-approximation}
    \tilde{A} = A_{\rm C}\times \hat{A}^{-1}\times A_{\rm R},
\end{equation}
where $A_{\rm C}$ and $A_{\rm R}$ are composed of $R$ columns and rows of $A$, respectively, and $\hat{A} \in \mathbb{R}^{R\times R}$ is a submatrix at their intersection. If $\hat{A}$ is a submatrix of maximal volume, {\it i.e.}, its determinant is maximal in modulus among all possible $R \times R$ submatrices of $A$, the maximal element $\hat{A}_{\max}$ of $\hat{A}$ provides a rather good approximation to the maximal element $A_{\max}$ of the whole matrix $A$~\cite{GOS10}. In practice, for a large matrix $A$ it is easier to find its submatrix of maximal volume rather than the element with the largest absolute value $A_{\max}$. 

\begin{figure}[h!]
\centering
\includegraphics[width=0.48\textwidth]{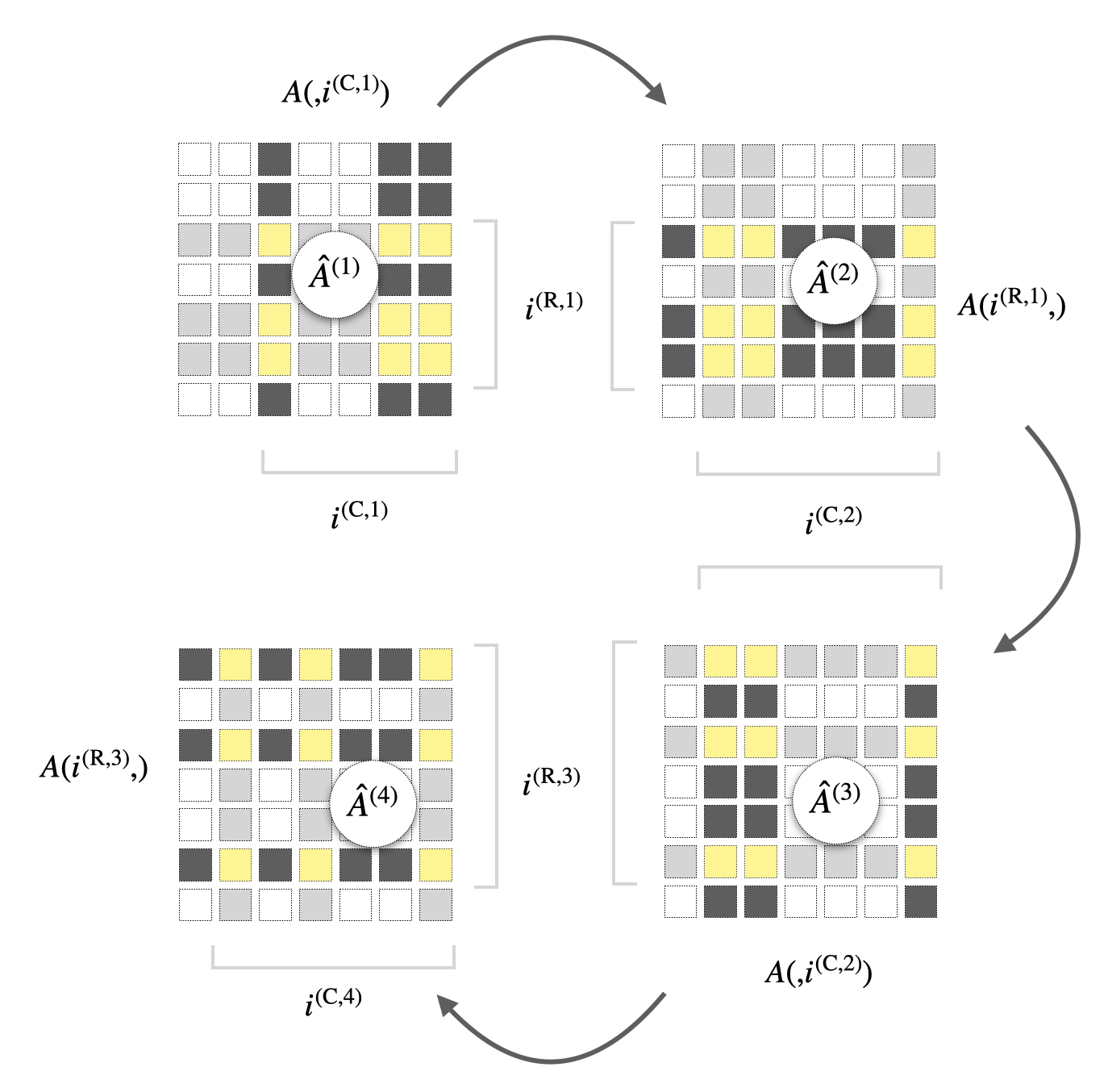}
\caption{A schematic of the TTOpt method for a two-dimensional function represented by the matrix. Dark gray bars show fixed rows or columns generated on the previous step, whereas bright gray bars mark rows or columns explored by the maxvol algorithm. On each iteration step, the TTOpt computes only a narrow submatrix of the original matrix $A$ for the selected rows or columns and identifies therein the maximal volume submatrix $\hat{A}^{(t)}$, as visualized by yellow squares at the intersection of dark gray rows (columns) and bright gray columns (rows). The maximum element of $A$ is searched in the generated series of submatrices $\lbrace \hat{A}^{(t)} \rbrace_{t=1}^T$.}
\label{fig:ttopt}
\end{figure} 

In two dimensions, the TTOpt method iteratively searches for maximal volume submatrices in the column and row spaces of matrix $A$, as schematically illustrated in Fig.~\ref{fig:ttopt}. This is done by means of the maxvol algorithm~\cite{GOS10} that allows to find the maximal volume submatrix $\hat B \in \mathbb{R}^{R\times R}$ of any rectangular matrix $B \in \mathbb{R}^{N\times R}$, where $N>R$ (for details, see~\cite{SCS22}). Starting from $R$ random columns of $A$ denoted by $i^{(\rm C,1)}$, the method allows one to estimate the maximal volume submatrix $\hat A^{(1)} \in \mathbb{R}^{R\times R}$ of the rectangular matrix $A(,i^{(\rm C,1)}) \in \mathbb{R}^{N_1\times R}$ and stores its row indices $i^{(\rm R,1)}$. Then, the maximal volume submatrix $\hat{A}^{(2)} \in \mathbb{R}^{R\times R}$ of the rectangular matrix $A(i^{(\rm R,1)},) \in \mathbb{R}^{R\times N_2}$ is obtained, and the corresponding column indices $i^{(\rm C,2)}$ are stored. Repeating this process for $T$ iterations, a series of {\it intersection} matrices $\lbrace \hat{A}^{(t)} \rbrace_{t=1}^T$ is produced, and the maximal element of the matrix $A$ is searched in these low-rank submatrices. 

\begin{figure*}[ht!]
\centering
\includegraphics[width=0.85\textwidth]{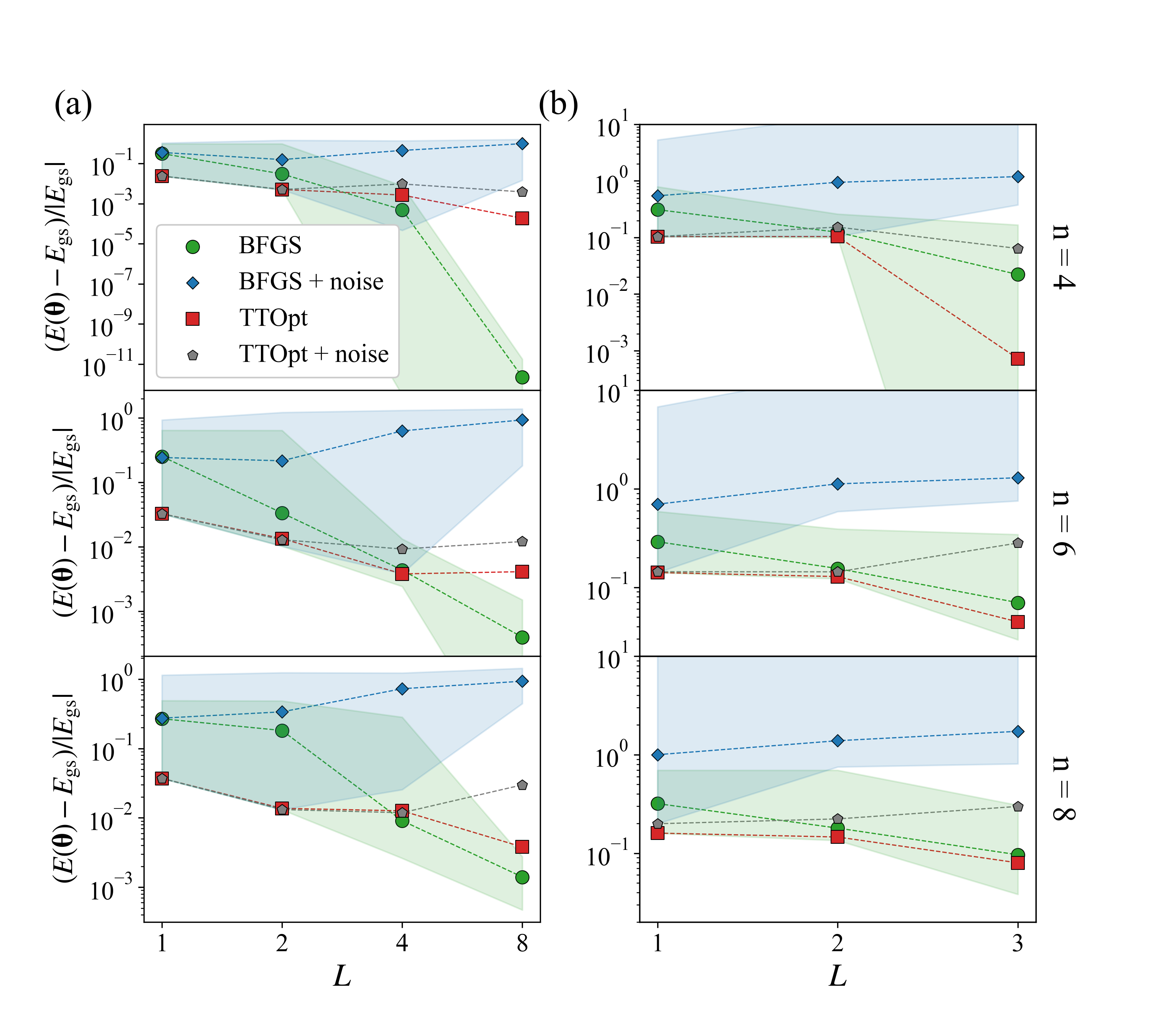}
\caption{Optimized cost function $E(\bm{\theta})$, as specified by the Hamiltonian~\eqref{TFIM}, relative to the exact ground-state energy $E_{\rm gs}$ plotted versus the ansatz depth $L$, where the optimization is performed as based on the BFGS and TTOpt optimizers. The VQE simulations are implemented for the TFIM~\eqref{TFIM} of $n=4$, 6, and 8 qubits under open boundary conditions with (a) HVA and (b) HEA being used as variational quantum circuits. The BFGS results are averaged over 100 random initial guess for the variational parameters $\bm{\theta}$. The green and blue shaded areas depict the dispersion of the optimized values for $E(\bm{\theta})$. The results with noise are obtained by applying the depolarizing quantum channel \eqref{eq:noise} for one- and two-qubit gates in quantum circuits, with the depolarizing parameter being equal to $\lambda = 0.005$.}
\label{fig:results}
\end{figure*} 

In the case of a multi-dimensional tensor $A = (A_{i_1i_2\ldots i_d}) \in~\mathbb{R}^{N_1\times N_2\times \ldots \times N_d}$, the direct implementation of the algorithm is more tricky. In this scenario, the TTOpt operates with the so-called unfolding matrices of the tensor $A$. The $k$-th unfolding $A_{k}$ of the tensor $A$ is the matrix $A_k \in \mathbb{R}^{(N_1 \ldots N_k) \times (N_{k+1}\ldots N_d)}$ with elements $A_k(\overline{i_1\ldots i_k}, \overline{i_{k+1}\ldots i_d}) = (A_{i_1i_2\ldots i_d})$ for all indices $i_k$. Starting from the first unfolding $A_1 \in \mathbb{R}^{N_1 \times (N_2 \ldots N_d)}$, the algorithm selects $R_1$ random columns $i_1^{(\rm C)}$ in $A_1$, calculates the maximal volume submatrix $\hat{A}^{(1)} \in \mathbb{R}^{R_1 \times R_1}$ of the rectangular matrix $A_1(,i^{(\rm C)}_1) \in \mathbb{R}^{N_1 \times R_1}$ and stores the resulting row indices $i^{(\rm R)}_1$. Then, the matrix $A_1(i^{(\rm R)}_1,) \in \mathbb{R}^{R_1 \times (N_2\ldots N_d)}$ is reshaped into the second unfolding $A_2 \in \mathbb{R}^{(R_1N_2) \times (N_3 \ldots N_d)}$ without explicitly evaluating its elements. Next, the method samples $R_2$ random columns $i^{(\rm C)}_2$ in $A_2$, calculates the maximal volume submatrix $\tilde{A}^{(2)} \in \mathbb{R}^{R_2 \times R_2}$ of the matrix $A_2(,i^{(\rm C)}_2) \in \mathbb{R}^{(R_1N_2) \times R_2}$ and stores the row indices $i^{(\rm R)}_2$. The algorithm then transforms the matrix $A_2(i^{\rm R}_2,) \in \mathbb{R}^{R_2 \times (N_3 \ldots N_d)}$ into the third unfolding going onwards.

The described operations, called sweeps, continue until the last dimension of the initial tensor $A$ is reached. After that, the process is repeated in the opposite direction, sampling now the row indices instead of the column ones. These sequences of forward and backward sweeps are continued until the algorithm converges or exceeds the number of requests to the objective function $A$. Similarly to the two-dimensional case, the maximal element after $T$ iterations is searched in the series $\lbrace \hat{A}^{(t)} \rbrace_{t=1}^T$. The complexity of the TTOpt method is 
\begin{equation*}
    \mathcal{O}(T d \times \max_{1\leq k \leq d}{(N_k R_k^3)}).
\end{equation*}

\noindent To speed up our VQE computations, we use the Quantum Tensor Train based (QTT) modification of the TTOpt. In this modification, the additional compression is applied based on quantization of the tensor modes~\cite{Ose10}. That is, if each dimension of the tensor $A$ is $N_k = p^q$, $k=1,\ldots,d$, with $p\geq 2$ and $q\geq 2$, the original tensor $A \in \mathbb{R}^{N_1 \times N_2 \times \ldots \times N_d}$ can be reshaped into the tensor $\tilde{A} \in \mathbb{R}^{p \times p \times \ldots \times p}$ of a higher dimensions $q\times d$, but with smaller modes of size $p$. The TTOpt is then applied to this modified {\it long} tensor $\tilde A$ instead of $A$. As pointed out in Ref.~\cite{SCS22}, this trick reduces the sizes of unfolding matrices, complexity, and execution time of the algorithm. For more details on the TTOpt implementation we refer the readers to Ref.~\cite{SCS22}.

{\it Numerical results.---}The VQE calculations are implemented using the Qiskit program suite~\cite{qiskit}. For minimizing the cost function, we make use of two different optimizers, including the gradient-based Broyden-Fletcher-Goldfarb-Shanno (BFGS) method~\cite{DS96} as implemented in Qiskit and derivative-free TTOpt~\cite{Chertkov}. For the BFGS optimizer, we restrict the maximum number of iterations to $10^5$ and average over 100 random initial guess for variational parameters. For the TTOpt, we set identical uniform grids on $[0,2\pi]$ of the size $2^8$ and apply the QTT-based modification of the optimizer with $p=2$ and $q=8$, where the maximum number of evaluations of the objective function is set to $5\cdot 10^6$.
For noisy simulations, we apply the depolarizing error channel \eqref{eq:noise} after all one- and two-qubit gates in a circuit with the depolarizing parameter equal to $\lambda = 0.005$, and we also restrict the TTOpt iterations to $10^5$ because of the reduced VQE performance in the presence of noise.

\begin{figure}[ht!]
\centering
\includegraphics[width=0.65\linewidth]{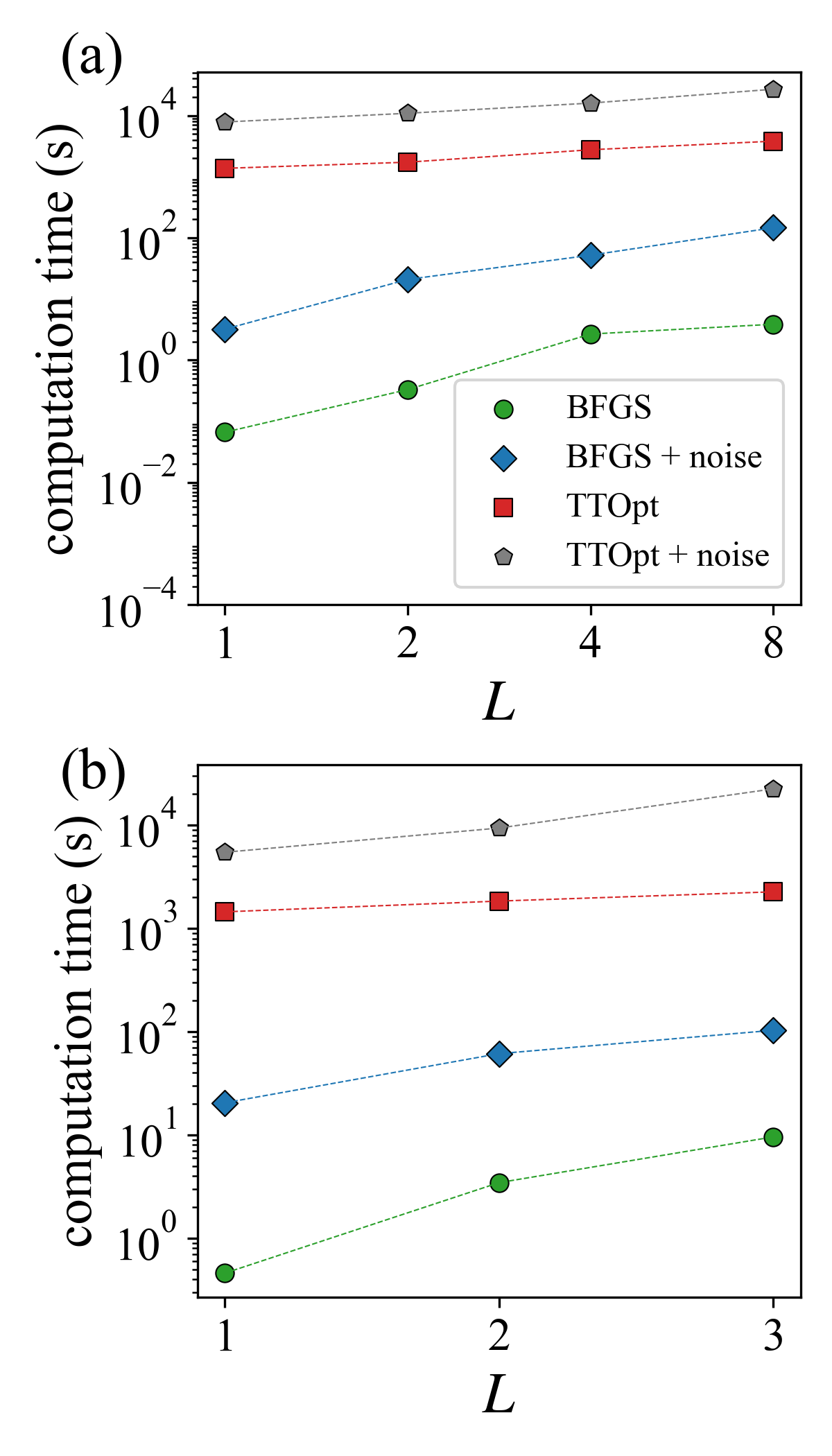}
\caption{VQE computation time (in s) for the TFIM \eqref{HVA-TFIM} of $n=4$ qubits when using two different optimizers --- BFGS and TTOpt --- upon varying the ansatz depth $L$, where simulations are performed with a) HVA and b) HEA being used as variational quantum circuits. The BFGS simulation time is averaged over 100 runs for random initial guess of the variational parameters.}
\label{fig:calc_time}
\end{figure} 

Our numerical findings are shown in Fig.~\ref{fig:results} for the TFIM~\eqref{TFIM} of $n=4$, 6, and $8$ qubits under open boundary conditions, where the VQE parametrized circuits are in the form of HVA and HEA of different depths $L$. A close inspection of Fig.~\ref{fig:results}(a) reveals that the HVA optimization with BFGS steadily improves with $L$ providing a proper approximation to the ground state 
starting from $L=4-8$ layers. However, for extremely shallow circuits down to $L=2$ layers, it is not robust to random initialization of variational parameters. As opposed, the TTOpt outperforms the BFGS optimizer
for $L=2-4$ layers, but its convergence improves with $L$ slower and can not compete in this regard to a gradient-based optimization for large~$L$.
This mainly conditioned by the number of variational parameters, the size of discretization grid in the TTOpt and the number of cost function evaluations.

In the presence of the depolarizing noise specified by Eq.~\eqref{eq:noise},
the situation is different. The BFGS optimizer completely fails in achieving convergence. As $L$ increases, the noise accumulates leading to a larger deviation between the ground state and optimized cost function values obtained by the BFGS optimizer in agreement with the results of Ref.~\cite{ZWC21}. 
On the contrary, the results of TTOpt do not change much with noise providing a reasonable accuracy in comparison to the BFGS optimizer. 
Thus, the TTOpt seems to be noise-resilient at least for the case of shallow circuits. 

Finally, if we switch to the HEA-type ansatz the results of TTOpt are even more impressive as illustrated in Fig.~\ref{fig:results}(b). 
The TTOpt outperforms the BFGS optimizer in the range of $L=1-3$ layers for both pure and noisy simulations. Meanwhile, the larger number of qubits to be involved the deeper circuits have to be utilized, making TTOpt more computationally demanding. To estimate the TTOpt performance, we compare VQE computation time for both BFGS and TTOpt optimizers in Fig.~\ref{fig:calc_time}. As one can see, the BFGS-based simulations are done in $0.1-10$~s, whereas the use of TTOpt requires more than $10^3$~s for success. The performance becomes even worse for noisy simulations as the TTOpt computation time increases to about $10^4$ s and grows exponentially with the circuit depth $L$ for both HVA and HEA variational quantum circuits. 

{\it Conclusion.---}In this work, we addressed a practically viable implementation of the derivative-free method for classical subroutine within variational quantum algorithms. It is based on the tensor train based optimization and allows one to effectively find extreme values of multi-dimensional tensors. We showed that this methodology is more robust to the depolarizing noise as opposed to gradient-based optimization techniques and provides a better convergence to the ground state in the case of extremely shallow circuits, being however more computationally heavy. We believe one can remedy the issue by a proper parallelization of the algorithm and leave this for future studies. 

{\it Acknowledgement.---}We acknowledge the support from the Russian Science Foundation Project 22-11-00074. We also acknowledge use of the computational resources at the Skoltech supercomputer ``Zhores''~\cite{Zhores} in our numerical simulations. 

\FloatBarrier

\bibliography{main.bbl}

\end{document}